\documentclass[a4paper,twocolumn,english,prl,superscriptaddress,showpacs,tight]{revtex4}
\usepackage{ae,aecompl}
\usepackage[T1]{fontenc}
\usepackage[latin1]{inputenc}
\usepackage{bm}
\usepackage{amsmath}
\usepackage{amssymb}
\usepackage{graphicx}
\usepackage{esint}

\makeatletter

\@ifundefined{textcolor}{}
{%
 \definecolor{BLACK}{gray}{0}
 \definecolor{WHITE}{gray}{1}
 \definecolor{RED}{rgb}{1,0,0}
 \definecolor{GREEN}{rgb}{0,1,0}
 \definecolor{BLUE}{rgb}{0,0,1}
 \definecolor{CYAN}{cmyk}{1,0,0,0}
 \definecolor{MAGENTA}{cmyk}{0,1,0,0}
 \definecolor{YELLOW}{cmyk}{0,0,1,0}
}

\usepackage{epsfig}
\usepackage{bm}
\usepackage{babel}

\makeatother

\usepackage{babel}

\begin{document}

\title{Confined Magneto-Optical Waves in Graphene}

\author{Aires Ferreira}

\affiliation{Graphene Research Centre and Department of Physics, National University
of Singapore, 2 Science Drive 3, Singapore 117542}

\affiliation{Department of Physics and Center of Physics, University of Minho,
P-4710-057, Braga, Portugal}

\author{N. M. R. Peres}

\affiliation{Graphene Research Centre and Department of Physics, National University
of Singapore, 2 Science Drive 3, Singapore 117542}

\affiliation{Department of Physics and Center of Physics, University of Minho,
P-4710-057, Braga, Portugal}

\author{A. H. Castro Neto}

\affiliation{Graphene Research Centre and Department of Physics, National University
of Singapore, 2 Science Drive 3, Singapore 117542}

\begin{abstract}
The electromagnetic mode spectrum of single-layer graphene subjected
to a quantizing magnetic field is computed taking into account intraband
and interband contributions to the magneto-optical conductivity. We
find that a sequence of weakly decaying \emph{quasi-}transverse-electric
modes, separated by magnetoplasmon polariton modes, emerge due to
the quantizing magnetic field. The characteristics of these modes
are tuneable, by changing the magnetic field or the Fermi energy. 
\end{abstract}


\maketitle

\section{Introduction\label{sec:Introduction}}

The field of plasmonics has been considered the photonics milestone
of the year 1998. To this choice contributed the landmark paper of
Ebbesen \emph{et al}. on the ``extraordinary optical transmission
through sub-wavelength hole arrays.''\cite{Ebbesen1998} The effect
was explained on the basis of the properties of surface-plasmon polaritons.\cite{PlasmonBook} 

Surface-plasmon polaritons (SPP) are electromagnetic surface waves,
guided by a metallic interface, resulting from the the coupling of
the electromagnetic field to the collective plasma excitations of
the metal.\cite{PlasmonBook} These guided modes are of importance
in fields as different as light guides at the nanoscale,\cite{Ashkan}
spectroscopy and sensing, enhancement of light absorption in solar
cells, enhanced Raman spectroscopy, and others.

Graphene, being an one-atom-thick metallic film, is an obvious candidate
for investigations on SPP. Recent research has established plasmon-based
enhanced Raman spectroscopy and photocurrent,\cite{Schedin2010,EchtermeyerPhoto}
as well as room-temperature prominent absorption peaks in the terahertz
spectral range,\cite{LongJuPlasmonics} and nanoscopy of mid-infrared
radiation confinement.\cite{dirac-plamons}

Both in the two-dimensional (2D) electron gas and in graphene, the
plasmon dispersion has a square-root dependence on the wave vector:
$\Omega_{\textrm{2D}}(\bm{q})\propto\sqrt{q}$. The linear dispersion
of the electrons in graphene, $\epsilon(\bm{q})=\pm v_{F}q$, where
$v_{F}$ is the Fermi velocity, implies that $\Omega_{\textrm{2D}}(\bm{q})\propto\sqrt{k_{F}q}$,
where $k_{F}\propto\sqrt{n_{e}}$ is the Fermi momentum and $n_{e}$
is the electronic density.\cite{rmp} If a grid of period $L$ is
superimposed on graphene, plasmons of wave number $q\sim1/L$ can
be excited.\cite{PlasmonBook,YuliyEPL} Furthermore, since $\Omega_{\textrm{2D}}(\bm{q})\propto\sqrt{k_{F}q}$,
we expect the scaling relation $\Omega_{\textrm{2D}}(\bm{q})\propto n_{e}^{1/4}L^{-1/2}$,
which has been observed experimentally.\cite{LongJuPlasmonics}

Graphene has a number of advantages over other metallic thin films
used in plasmonics; e.g., the ability of changing its carrier concentration
using a gate, allowing a fine control over the frequency range for
plasmonic excitations,\cite{dirac-plamons} and long propagation lengths
as compared to conventional SPP.\cite{plasmonics_graph_zerofield_Jablan,Koppens}
Furthermore, inhomogeneous doping in a single graphene sheet allows
the drawing of SPP propagation paths.\cite{Ashkan}

When an external magnetic field perpendicular to a 2D electron gas
is applied, hybridization between cyclotron excitations and plasmons
occurs, originating magnetoplasmon modes.\cite{Chiu_Quinn_1974,exp2dmspp,Bychkov2008,Berman2008}
The presence of the magnetic field gives rise to strong absorption
peaks, making the dispersion of electromagnetic modes very sensitive
to the frequency. In this work, we demonstrate that in addition to
magnetoplasmon polaritons (MPP), graphene in a magnetic field supports
extremely weakly damped modes, which due to their resemblance to conventional
transverse electric modes, are here referred to as \emph{quasi}-transverse-electric
(QTE). This paper is organized as follows: In Sec.~\ref{sec:Dispersion_Relation},
we overview the dispersion relation of electromagnetic modes supported
by 2D electron systems. We revisit the simpler problem of zero external
field, for which two types of modes can exist: SPP and weakly damped
modes (transverse electric) with characteristics similar to photons
propagating in a dielectric. The magneto-optical response of graphene
in the presence of a quantizing magnetic field is described in Sec.~\ref{sec:Magneto-Optical-modes}.
The full mode dispersion (MPP and QTE modes) is calculated in the
presence of disorder by employing the optical limit approximation
to the conductivity. The losses, confinement, and polarization of
the solutions are studied carefully. The outlook and conclusions are
presented in Sec.~\ref{sec:Outlook-and-Conclusions}. Finally, technical
details and derivations are given in appendices.

\section{dispersion relation\label{sec:Dispersion_Relation}}

We consider an infinite graphene film in the $xy$ plane embedded
in a dielectric medium of permittivity (permeability) $\epsilon$
($\mu$).\cite{comment_2dielectrics} A static quantizing magnetic
field is applied along the transverse ($z$) direction. We focus on
electromagnetic modes propagating along the $x$ axis, 
\begin{equation}
\boldsymbol{E}(\boldsymbol{r},t)=\boldsymbol{E}_{0}e^{i(qx-\Omega t)}e^{-\kappa|z|}\,.\label{eq:plasmonic_sol_em}
\end{equation}
The symbols have the usual meanings: $\Omega$ denotes the angular
frequency, $q$ is the complex longitudinal wave vector, and $\kappa$
encodes the amount of confinement along the transverse direction.
Maxwell equations relate these quantities according to the general
relation $q^{2}=\kappa^{2}+\epsilon\mu\Omega^{2}$, so that in general
both $q$ and $\kappa$ are complex quantities. Note that a similar
equation holds for the magnetic field $\boldsymbol{H}(\boldsymbol{r},t)$.
Throughout, we employ SI units and the notation $z=z^{\prime}+iz^{\prime\prime}$
for complex variables.

The dispersion relation of electromagnetic modes follows from the
boundary conditions for the fields at the interface $z=0$ (Ref.~\onlinecite{Chiu_Quinn_1974}):
\begin{equation}
\textrm{det}\,\left(\begin{array}{cc}
\frac{i\kappa\sigma_{L}(q,\Omega,B)}{2\epsilon\Omega}+1 & \frac{{\cal Z}}{2}\sigma_{H}(q,\Omega,B)\\
\frac{{\cal Z}}{2}\sigma_{H}(q,\Omega,B) & \frac{i\mu\Omega\sigma_{L}(q,\Omega,B)}{2\kappa}-1
\end{array}\right)=0\,,\label{eq:fundamental_relation}
\end{equation}
where $B$ is the intensity of the magnetic field, ${\cal Z}=\sqrt{\mu/\epsilon}$
is the impedance of the surrounding medium and $\sigma_{L}$ ($\sigma_{H}$)
denotes the longitudinal (Hall) conductivity of graphene. The physical
solutions of the above equation, $q=q(\Omega)$, contain the full
mode spectrum of the system---a derivation of the dispersion relation
is given in Appendix~A. The characteristics of the mode spectrum
are determined by the conductivity tensor. The latter depends on $q$,
$\Omega$, $B$, and, generally, also on the Fermi energy $E_{F}$,
temperature and sample-specific broadening parameters. In the present
work, we neglect the dependence of $\sigma_{L(H)}$ on the in-plane
wave vector $q$, and thus, hereafter, $\sigma_{H(L)}$ denotes the
optical (local) limit of the dynamical conductivity, i.e.,~$\sigma_{H(L)}\equiv\sigma_{H(L)}(\Omega,B)=\sigma_{H(L)}(0,\Omega,B)$.
The latter is justified for small wave vectors, more precisely for
$ql_{B}\ll1$, where $l_{B}=\sqrt{\hbar/eB}$ denotes the magnetic
length ($-e<0$ is the electron charge). 

In the absence of an external field $B$, the solutions of Eq.~(\ref{eq:fundamental_relation})
are the so-called transverse electric and transverse magnetic modes,
namely, $\kappa=i\Omega\mu\sigma_{L}(\Omega,0)/2$ and $\kappa=2i\Omega\epsilon/\sigma_{L}(\Omega,0)$,
respectively. In the former, the electric field is perpendicular to
the direction of propagation, the mode dispersion is close to the
light line, $q^{\prime}(\Omega)\simeq\Omega/c$, and damping is small
($q^{\prime\prime},\kappa^{\prime}\ll q^{\prime}$). Transverse electric
modes require a negative reactive conductivity, $\sigma_{L}^{\prime\prime}(\Omega,0)<0$,
and hence are not observed in conventional 2D gases (we note that
when $\sigma_{L}^{\prime\prime}<0$ and $\sigma_{L}^{\prime}/\vert\sigma_{L}^{\prime\prime}\vert\ll1$,
the behavior of the system resembles that of a dielectric). On the
other hand, transverse magnetic waves are confined to the metallic
surface, featuring large-field localization, thus having important
applications in sub-wavelength optics and plasmonics.\cite{Barnes2010}
In graphene, both modes can exist because near the interband threshold,
$\Omega=2E_{F}/\hbar$, the function $\sigma_{L}^{\prime\prime}(\Omega,0)$
changes sign.\cite{ZieglerPRL} Coupling and guiding of transverse
electric modes by graphene in zero field have been recently reported
in Ref.~\onlinecite{TEpolarizer}. 

\begin{figure}
\begin{centering}
\includegraphics[clip,width=0.6\columnwidth]{fig_01a}\includegraphics[clip,width=0.4\columnwidth]{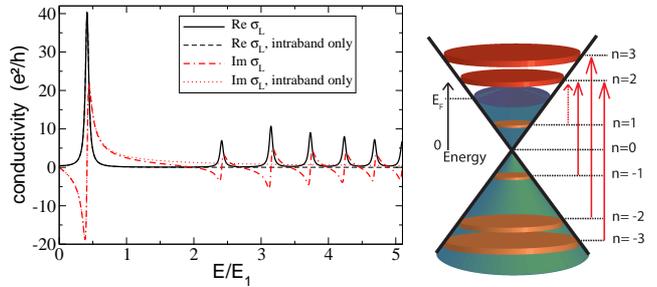} 
\par\end{centering}

\caption{\label{fig:01} Doped graphene in a magnetic field at zero temperature
($N_{F}=1$). \textbf{Left}: $\sigma_{L}$ is plotted as a function
of $E/E_{1}$, with $E=\hbar\Omega$ and $E_{1}=E_{1}(B)$. Clearly,
$\sigma_{L}^{\prime\prime}<0$ in various energy intervals. The full
quantum calculation is seen to be crucial for energies above $\sim2E_{1}$.
We have taken $\Gamma=0.03E_{1}(B)$, which, for $B$ in the range
1-10~T, is consistent with experimental values of $\Gamma$ (see
text). \textbf{Right}: The first few electronic transitions are shown.
The Fermi energy lies slightly above the Landau level with $n=1$,
and thus only transitions to levels with $n\ge2$ are allowed. The
(small) dashed arrow stands for the intraband transition responsible
for the strong peak observed near at $\left(\sqrt{2}-1\right)E_{1}\sim0.4E_{1}$.
Other arrows represent interband transitions.}
\end{figure}

\section{Magneto-Optical modes\label{sec:Magneto-Optical-modes}}

When a magnetic field is turned on, electrons acquire considerable
cyclotronic energies via the Lorentz force, and at sufficiently high
fields, the continuum Dirac quasi-particle spectrum condensates into
degenerated Landau levels (LLs) {[}see Fig.~\ref{fig:01} (right
panel){]}, with energies given by $E_{n}(B)=s_{n}\sqrt{2|n|}\hbar v_{F}/l_{B}$;
here, $n$ is the LL index ($n=0,\pm1,\pm2,...$), $s_{n}\equiv\textrm{sign}(n)$,
and $v_{F}\simeq10^{6}$~m/s denotes the Fermi velocity of carriers
in graphene. 

We would like to investigate how the zero-field mode spectrum changes
due to the quantizing magnetic field. To this end, we employ linear-response
theory within the Dirac cone approximation to obtain an expression
for $\sigma_{L(H)}(\Omega,B)$ with both intraband and interband contributions
included (see Appendix B). In order to account for disorder, we have
used an energy-independent LL broadening $\Gamma$ with ratios $\Gamma/E_{1}$
consistent with the values 1-10~meV found in pump-probe experiments
performed in epitaxial and exfoliated graphene samples,\cite{pump_probe_epitaxial,pump_probe_exfoliated}
and on infrared spectroscopy studies of the Drude conductivity of
graphene.\cite{Horgn2011} The renormalization of the optical conductivity
due to the electron-phonon interactions is neglected. From the theoretical
studies, taking into account the $E{}_{2g}$ optical mode at 200~meV,\cite{Carbotte_Phonons,Carbotte_Phonons_2}
we expect the latter approximation to be valid for frequencies below
that of the optical phonon branch.

The features of $\sigma_{L}(\Omega,B)$ for doped graphene in a quantizing
field are determined by the amount of disorder and the LL occupancy
number of the graphene sample. The latter is defined as $N_{F}=\textrm{int}[(E_{F}/E_{1}(B)){}^{2}]\ge0$
and yields the number of occupied (empty) electron- (hole-) degenerate
LLs for $E_{F}>0$ ($E_{F}<0$). In Fig.~\ref{fig:01} (left panel),
we plot $\sigma_{L}(\Omega,B)$ as a function of energy $\hbar\Omega$.
These curves have $N_{F}=1$, thus covering a wide-range of $E_{F}$
and $B$ values. (For concreteness, throughout, our plots refer to
systems with $N_{F}=1$, except for one occasion. We complement the
discussion with analytic expressions that can be used to compute the
several quantities for arbitrary $N_{F}$.) The magneto-optical conductivity
is seen to consist of an intraband term with spectral weight located
at the lower end of the spectrum {[}the strong peak located at $\hbar\omega\simeq(\sqrt{N_{F}+1}-\sqrt{N_{F}})E_{1}${]}
and interband high-frequency contributions originating a series of
peaks above the interband threshold, $E\simeq2\sqrt{N_{F}}E_{1}$.
The number of peaks depends on $N_{F}$ (e.g.,~for $N_{F}=2$, the
first interband peak seen in Fig.~\ref{fig:01} is suppressed due
to Pauli blocking) and their shape depends also on $\Gamma$. For
a comprehensive discussion of the magneto-optical response of graphene,
see Refs.~\onlinecite{Sharapov,EOM}. 

\begin{figure}
\begin{centering}
\includegraphics[clip,width=0.95\columnwidth]{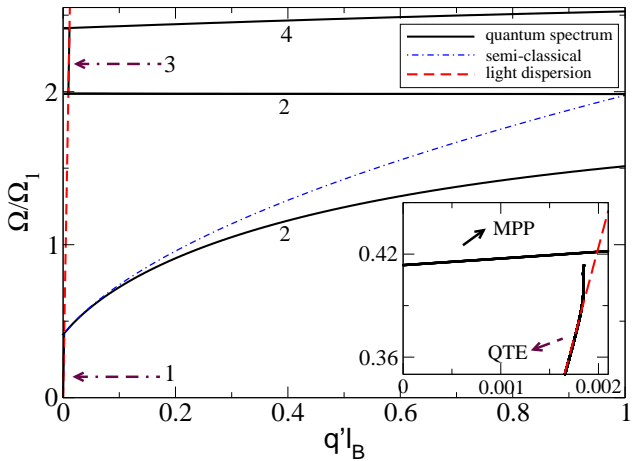}
\par\end{centering}

\caption{\label{fig:02}Mode frequency $\Omega$ for graphene in vacuum is
plotted as a function of the wave vector $q^{\prime}$ for doped graphene
{[}solid (black) line{]}. The wave vector $q^{\prime}$ is given in
units of the inverse of the magnetic length. For completeness, semi-classical
solution {[}dashed-dotted (blue) line{]} and the light dispersion
{[}dashed (red) line{]} are shown. Inset: Mode spectrum near at QTE-MPP
transition at $\Omega\simeq0.4\Omega_{1}$. The numbers in the main
panel identify the distinct branches up to $\Omega\simeq2.5\Omega_{1}$.
For consistency with the local approximation to the conductivity,
we have restricted our plots to the region with $q^{\prime}(\Omega)l_{B}\le1$.
A considerable modification to the spectrum is expected for $q^{\prime}\sim l_{B}^{-1}$
due to non-local effects in the dynamical conductivity (see text).
Other parameters as in Fig.~\ref{fig:01}. }
\end{figure}

In 2D electron gases, plasmons and cyclotronic excitations (with frequency
$\omega_{\textrm{c}}$) hybridize leading to the well-known semi-classical
magnetoplasmon spectrum, $\Omega^{2}=\Omega_{\textrm{2D}}^{2}+\omega_{\textrm{c}}^{2}$.\cite{Chiu_Quinn_1974,exp2dmspp,Goerbig}
In view of the strong contribution of interband transitions to $\sigma_{L}^{\prime\prime}$
(e.g.,~see the discrepancy between the semi-classical calculation
and the quantum formula, even at low frequencies, in the left panel
of Fig.~\ref{fig:01}), this formula should be of limited applicability
in graphene. Moreover, Shubnikov\textendash{}de Haas oscillations
originate many frequency regions with $\sigma_{L}^{\prime\prime}<0$,
and hence, similarly to the 2D magnetized electron Fermi gas,\cite{Chiu_Quinn_1974,Bardos_Frankel_1994}
we may expect the splitting of the mode spectrum into many branches. 

\begin{figure}
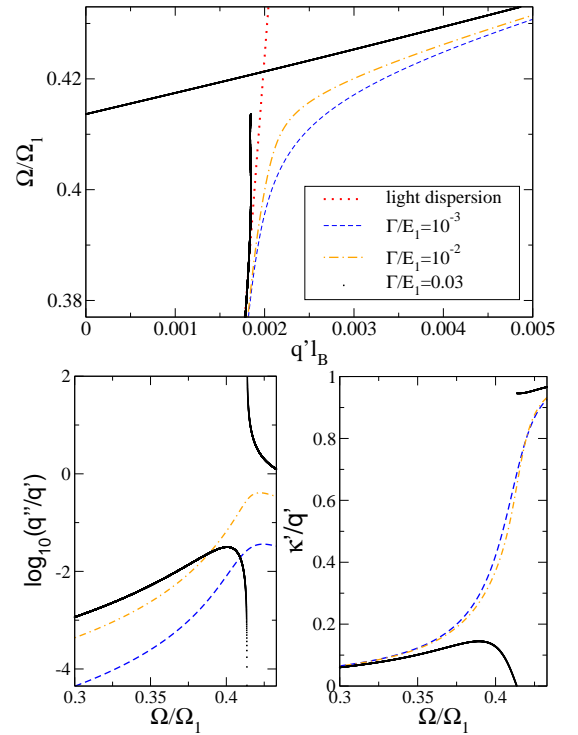

\begin{centering}
\includegraphics[clip,width=0.82\columnwidth]{fig_03a}
\par\end{centering}

\centering{}\includegraphics[clip,width=0.8\columnwidth]{fig_03b}\caption{\label{fig:03}Mode spectrum and decay properties near a QTE-MPP transition.
Top panel: The dispersion relation is plotted near at $\Omega\simeq0.4\Omega_{1}$
for several broadening values (indicated in the legend). Bottom panel:
The ratios $q^{\prime\prime}(\Omega)/q^{\prime}(\Omega)$ and $\kappa^{\prime}(\Omega)/q^{\prime}(\Omega)$
are shown for the same parameter region in the top panel. For ease
of visualization, the range for the vertical axis in left panel is
limited. Other parameters as in Fig.~\ref{fig:02}.}
\end{figure}

The optical-limit solution of the dispersion relation for graphene
in vacuum in the presence of the magnetic field is given in Appendix
C {[}Eq.~(\ref{eq:sol_mag1}){]}. The respective mode spectrum is
shown in Fig.~\ref{fig:02}, with frequency given units of the LL
energy scale $\Omega_{1}\equiv E_{1}/\hbar$. Note that $q^{\prime}=q^{\prime}(\Omega)$
has been plotted in the horizontal axis which helps visualizing the
dispersion relation. Only non-exponentially growing solutions $\kappa^{\prime}(\Omega)>0$
are shown. The most notorious feature is the existence of a series
of well-defined branches, labeled by the integers $n=1,2,...,$ etc.
These branches can be divided into two distinct sets, namely, the
set of branches with dispersion close to the light line ($n$ odd)
and the remaining ($n$ even). The former will be shown to have the
basic properties of transverse electric modes and hence are termed
QTE, whereas the latter are MPP modes (with polarization not necessarily
similar to transverse magnetic modes). 

Our results borne out two peculiar features of graphene: i) $n$ even
branches have two distinct solutions for each wave vector $q^{\prime}$.
This degeneracy is a result of hybridization between even and odd
modes, and ii) the frequency domain size of each branch is non-uniform
due to the structure of LLs in graphene. The first branch occupies
a region $[0,\tilde{\Omega}_{1}[$, the second $[\tilde{\Omega}_{1},\tilde{\Omega}_{2}[$,
etc., where $\tilde{\Omega}_{n}$ is defined to be the $n$-th node
of the reactive longitudinal conductivity, $\sigma_{L}^{\prime\prime}(\tilde{\Omega}_{n},B)=0$.
For the system under discussion, the first two nodes read $\tilde{\Omega}_{1}\simeq0.4\Omega_{1}$
and $\tilde{\Omega}_{2}\simeq2\Omega_{1}$ (see Fig.~\ref{fig:01}). 

Let us now discuss with detail the intraband region ($\Omega\lesssim\Omega_{1}$)
spanning two branches, $n=1$ and $n=2$. Here, the magneto-optical
transport is predominantly semi-classical, and hence the relevant
frequency scale is the cyclotron frequency, $\omega_{\textrm{c}}=ev_{F}^{2}B/|E_{F}|$.
For $\tilde{\Omega}_{1}>\Omega>0$, the dispersion curve is pinned
to the light dispersion line, $q^{\prime}(\Omega)\simeq\Omega/c$,
except for frequencies approaching $\tilde{\Omega}_{1}$ (Fig.~\ref{fig:02}).
The detachment of the QTE mode from the light line signals the onset
of a rapid increase of $\sigma_{L}^{\prime\prime}$, as a result of
an absorption peak in the vicinity of $\omega_{\textrm{c}}$. A simple
formula for the transition frequency $\tilde{\Omega}_{1}$ can be
obtained by approximating $\sigma_{L}$ by its semi-classical value\cite{comment2}
\begin{equation}
\tilde{\Omega}_{1}\approx\sqrt{\omega_{\textrm{c}}{}^{2}-\Gamma^{2}/\hbar^{2}}\,.\label{eq:Omega_Zero}
\end{equation}
Using the above expression, we obtain $\tilde{\Omega}_{1}\approx0.4\Omega_{1}$
in good agreement with the exact numerical results.

The properties of the electromagnetic modes are specially sensitive
to electronic disorder close to the frequencies at which the transitions
occur, since the conductivity is strongly dependent on $\Gamma$ in
the vicinity of $\Omega=\tilde{\Omega}_{n}$. A closer look to the
region with $\Omega\sim\tilde{\Omega}_{1}$ is provided in Fig.~\ref{fig:03}
for several values of $\Gamma$. This figure shows that for small
broadening the transition at $\tilde{\Omega}_{1}$ can be continuous.
This is further elucidated in the lower panel of Fig.~\ref{fig:03},
containing a study of $\kappa^{\prime}(\Omega)$ and $q^{\prime\prime}(\Omega)$;
these quantities measure the 2D confinement and longitudinal losses,
respectively. For the smallest values of $\Gamma$ considered, we
clearly observe a smooth transition from a weakly decaying mode (typical
of transverse electric modes) to a confined mode with considerable
losses (typical of SPP and MPP). A remark is in order: near at $\tilde{\Omega}_{1}$,
the $n=1$ (QTE) mode for $\Gamma=0.03E_{1}$ clearly displays superluminal
group velocities. The latter is a manifestation of anomalous dispersion,
for which the concept of group velocity no longer describes signal
propagation.\cite{Jackson} We believe that the velocity of signal
propagation in the anomalous region equals its upper bound value,
$c$, given that the mode is essentially undamped ($q^{\prime\prime}$
and $\kappa^{\prime}$ reaching $10^{-5}q^{\prime}$).

The simultaneous presence of the two distinct branches ($n=1,2$)
in the intraband region ($\Omega\lesssim\Omega_{1}$) for a given
broadening value can only occur for sufficiently high fields, $B>B_{c}\equiv|E_{F}|\Gamma/e\hbar v_{F}^{2}$,
otherwise one obtains a single type of solution with plasmon character,
$\Omega\sim\sqrt{q^{\prime}}$, as for $B=0$.\cite{Stern_Polarizability2D,Shung_Graphite}
The reason is that for $B<B_{c}$, the reactive part of the semi-classical
longitudinal conductivity is always positive below the interband threshold,
thus forbidding the existence of QTE modes.

The semi-classical expression for the MPP dispersion $\Omega_{\textrm{(MPP})}(q^{\prime})$
can be derived assuming $T=0$ and ignoring the interband contributions
to the magneto-optical conductivity, 
\begin{equation}
\Omega_{\textrm{(MPP})}(q^{\prime})\simeq\sqrt{[\Omega_{\textrm{2D}}(q^{\prime})]^{2}+\omega_{\textrm{c}}^{2}}\,,\label{eq:omega_MPP_semiclassical}
\end{equation}
where $\Omega_{\textrm{2D}}(q^{\prime})=(e/\hbar)\sqrt{q^{\prime}|E_{F}|/(2\pi\epsilon)}$
is the graphene's plasmon-polariton dispersion in zero field. The
above expression is valid for $q^{\prime}\gg\Omega/c$, and for $\Omega$
within the regime of validity of the semi-classical transport theory,
$\Omega\ll2E_{F}/\hbar$ (see dotted-dashed line in the main panel
of Fig.~\ref{fig:02}). Equation.~(\ref{eq:omega_MPP_semiclassical})
predicts an increase of MPP's frequency due to the presence of a magnetic
field. A derivation of the above formula is given in Appendix C. This
result coincides with the semi-classical magnetoplasmon spectrum for
a 2D electron gas\cite{Chiu_Quinn_1974,exp2dmspp} and it has been
obtained in Ref.~\onlinecite{Goerbig} by studying the polarizability
of graphene in a magnetic field.

\begin{figure}
\centering{}\includegraphics[clip,width=0.9\columnwidth]{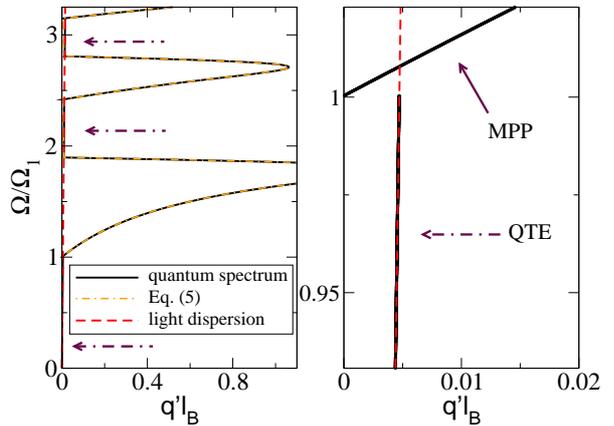}\caption{\label{fig:04}Mode spectrum for undoped graphene in a magnetic field
($N_{F}=0$) {[}solid (black) line{]}. We have taken $\Gamma=0.12E_{1}$
in this figure. The light dispersion is shown in the dashed (red)
line. Arrows in the left panel indicate the first three QTE modes. }
\end{figure}

We now turn our attention to the high-frequency part of the spectrum,
where new branches ($n=3,4,...$) emerge due to interband transitions.
Figure~\ref{fig:02} shows that the quantum calculation (solid line)
deviates considerably from the semi-classical result already at $\Omega\approx\Omega_{1}$.
In particular, the quantum corrections cause a considerable slow down
of the MPP's group velocity, $v_{\textrm{g}}=[dq^{\prime}/d\Omega]^{-1}$,
relative to its semi-classical value. This effect comes from the superposition
of interband resonances tails that contribute with substantial weight
even well below the interband threshold. For instance, in the range
$\Omega/\Omega_{1}\approx[1,1.5]$, the interband terms yield a correction
to the conductivity of about $\sim0.5i\, e^{2}/h$ (see Fig.~\ref{fig:01})
explaining the bending of the solid curve relatively to the dashed-dotted
curve in Fig.~\ref{fig:02}. Near at $\Omega=\tilde{\Omega}_{2}\approx2\Omega_{1}$,
$\sigma_{L}^{\prime\prime}$ changes sign again, and a large-bandwidth
($\approx0.5\Omega_{1}$) QTE mode develops. The first two QTE modes
are indicated by arrows in Fig.~\ref{fig:02}. Their dispersion relation
is well approximated by $q^{\prime}(\Omega)\simeq\Omega/c$, except
within the QTE/MPP crossovers ($\Omega\approx\tilde{\Omega}_{n}$,
with $n$ odd), where $q^{\prime}(\Omega)$ acquires a complex form
due the strong variation of the optical properties induced by sharp
absorption peaks (see discussion above). 

The full dispersion relation for the MPP branches is rather cumbersome
because, as noted above, away from the semi-classical region, many
interband terms contribute to the spectral weight around a particular
frequency; see Eq.~(\ref{eq:conductivity}) and text therein. A compact
expression for $q(\omega)$ valid for every MPP branch can still be
obtained by considering $T=0$ and neglecting the Hall conductivity
term in Eq.~(\ref{eq:fundamental_relation}). These approximations
are justified since (i) for a quantizing magnetic field, the conductivity
of graphene does not vary significantly with temperature, and (ii)
$\sigma_{H}$ can be shown to provide a small correction only in the
vicinity of each $\tilde{\Omega}_{n}$. We obtain,
\begin{align}
\sqrt{q^{2}-\epsilon\mu\Omega^{2}} & \simeq\frac{i\epsilon\Omega h}{e^{2}}\left(\frac{\Gamma/E_{1}(B)}{1-i\hbar\Omega/\Gamma}\right)\Psi^{-1}(\Omega,B)\,,\label{eq:interband_spectrum}\\
\Psi(\Omega,B) & =\sum_{n=|N_{F}|}^{N_{\textrm{cut}}\,\prime}\frac{E_{1}(B)/\Delta_{n}(B)}{(1-i\hbar\Omega/\Gamma)^{2}+\Delta_{n}(B)^{2}/\Gamma^{2}}\,.\label{eq:auxiliary}
\end{align}
In the above, $\Delta_{n}(B)$ stands for the $n$-th interband resonance
energy, defined as $\Delta_{n}(B)\equiv E_{n+1}(B)+E_{n}(B)$. The
prime in the summation sign indicates that if $N_{F}\neq0$, the first
term is to be halved. Also, a cutoff $n\le N_{\textrm{cut}}$ must
be taken when computing this summation (see Appendix B). For simplicity,
the above expression for $\Psi(\Omega,B)$ only includes the interband
contribution to $\sigma_{L}$. The inclusion of the intraband spectral
weight {[}see Eq.~(\ref{eq:sigma_L_intra}){]} is straightforward
and plays a role only in the first MPP branch. Figure~\ref{fig:04}
shows that the MPP spectrum computed from Eq.~(\ref{eq:interband_spectrum})
can not be distinguished from the full calculation. The latter agreement
extends down to low frequency (the first MPP branch) because the system
plotted in Fig.~\ref{fig:04} is half-filled (i.e.,~$N_{F}=0$).
Indeed, according to our definition of interband and intraband contributions
to the conductivity (see Appendix B), $\Psi(\Omega,B)$ already contains
the full spectral weight.

We have discussed the general features of the mode spectrum of graphene
under a quantizing electromagnetic field. It has been shown to consist
of several branches, with two possible types of modes. In what follows,
we demonstrate that MPP modes have conventional decaying properties
of SPP, whereas the QTE modes are essentially non-decaying, with electric
field nearly transverse (hence their name). The MPP solutions will
be shown to have a rich polarization diagram without a clear transverse
magnetic character.

\subsection*{Decaying and polarization properties}

\begin{figure}
\centering{}\includegraphics[width=0.9\columnwidth]{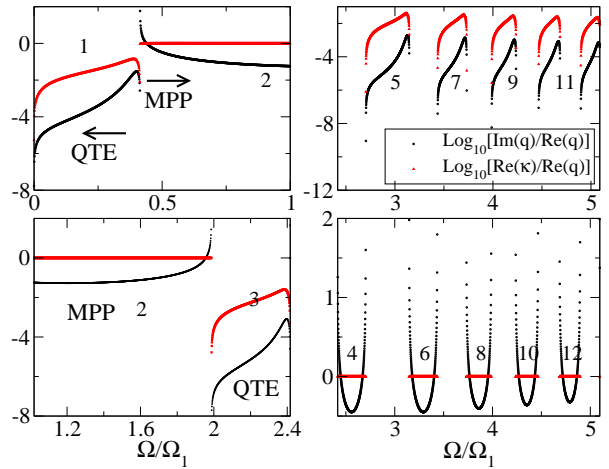}\caption{\label{fig:05}The ratios $q^{\prime\prime}/q^{\prime}$ and $\kappa^{\prime}/q^{\prime}$,
which characterize the longitudinal and transverse decaying rates,
respectively, are plotted in logarithmic scale as a function of the
mode frequency. Several panels are given to help in visualizing the
difference between QTE and MPP modes. (System parameters as in Fig.~\ref{fig:01}.)}
\end{figure}

The decaying properties of the modes are summarized in Fig.~\ref{fig:05}.
QTE modes display large localization length in $x$ ($z$) direction,
namely, $q^{\prime\prime}/q^{\prime}$ ($\kappa^{\prime}/q^{\prime}$)
in the range $10^{-7}$-$10^{-2}$ ($10^{-5}$-$10^{-1}$) (see for
instance the first branch; top left panel). MPP modes, on the other
hand, always show considerable decay along the $z$ direction, $\kappa^{\prime}=\mathcal{O}(q^{\prime})$,
indicating strong confinement. The losses in the propagation direction
$x$, on the other hand, vary appreciably and are determined by the
graphene's absorption at the specific MPP frequency; e.g.,~in the
range $0.4\Omega_{1}$-$1.0\Omega_{1}$, the longitudinal decay rate
$q^{\prime\prime}$ varies in the range 0.1$q^{\prime}$-100$q^{\prime}$,
with maximum loss occurring near at the cyclotron frequency ($\simeq0.4\Omega_{1}$),
where the MPP lies to the left of the light line (see top panel in
Fig.~\ref{fig:03}).

Although the losses and confinement reported here have orders of magnitude
comparable to those in the absence of a magnetic field\cite{plasmonics_graph_zerofield_Jablan},
the strong dependence of these quantities on the frequency is exclusive
to the 2D interface subjected to a strong external magnetic field
(Fig.~\ref{fig:05} shows that the decay characteristics can vary
by several orders of magnitude around at $\Omega=\tilde{\Omega}_{n}$
for all $n$.) An important effect of the magnetic field is to allow
for QTE modes with lower losses than the zero-field transverse electric
mode in specific frequency intervals; for instance, above $2.5\Omega_{1}$,
$q^{\prime\prime}/q^{\prime}$ can reach a minimum value of the order
of $10^{-7}$ , whereas for $B=0$ (and $E_{F}\sim0.1$~eV) its ratio
is about $\sim10^{-5}$.

\begin{figure}
\centering{}\includegraphics[clip,width=1\columnwidth]{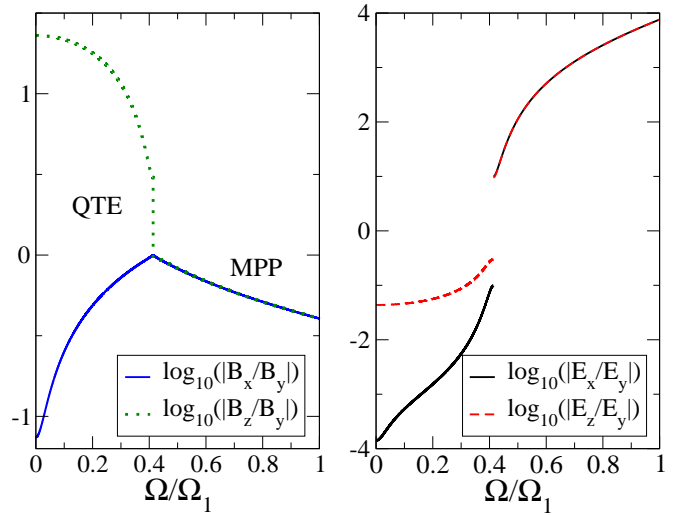}\caption{\label{fig:06}Polarization properties of the electromagnetic modes
within the first two branches. The left panel shows the complex modulus
of the magnetic ratios $\mathcal{B}_{xy}$ and $\mathcal{B}_{zy}$,
whereas the respective electric counterparts, $\mathcal{E}_{xy}$
and $\mathcal{E}_{zy}$, are given in the right panel. (System parameters
as in Fig.~\ref{fig:01}.)}
\end{figure}

In order to complete our study, we demonstrate that the electric field
of QTE modes are essentially transverse and study how the longitudinal
(transverse magnetic) character of MPP modes depend on the wave vector.
To this end, we compute the ratios $\mathcal{E}_{xy(zy)}\equiv E_{x(z)}/E_{y}$
and $\mathcal{B}_{xy(zy)}\equiv B_{x(z)}/B_{y}$. Combining Maxwell
equations and the dispersion relation, Eq.~(\ref{eq:fundamental_relation}),
we easily obtain, 
\begin{align}
\mathcal{E}_{xy} & =\left(\sigma_{L}-\frac{1}{i\omega}\frac{2\kappa}{\mu}\right)/\sigma_{H}\,,\label{eq:exy}\\
\mathcal{B}_{xy} & =i\frac{\kappa}{\omega}\left(2-\frac{\kappa}{i\omega\epsilon}\sigma_{L}\right)/\mu\sigma_{H}\,,\label{eq:bxy}
\end{align}
and $\mathcal{E}(\mathcal{B})_{zy}=[\text{sign}(z)iq/\kappa]\mathcal{E}(\mathcal{B})_{xy}$,
where it is assumed $B>0$ so that $\sigma_{H}\neq0$. These quantities
are plotted in Fig.~\ref{fig:06} for a frequency range spanning
the $n=1$ and $n=2$ branches. Below $0.4\Omega_{1}$, the electric
components ratios $\mathcal{E}_{xy(zy)}$, plotted in the right panel,
are found to have magnitude in the range $10^{-3}$-$10^{-2}$ ($10^{-2}$-$10^{-1}$),
confirming that the electric field of QTE modes lies prominently along
the $y$ axis, resembling pure transverse electric modes (which have
$\mathcal{E}_{xy(zy)}=0$). Similar conclusions can be drawn for the
remaining QTE branches.

The polarization of MPP modes is found to lack a clear longitudinal
character. Let us focus the branch $n=2$. For $\Omega\lesssim\Omega_{1}$
(see left panel of Fig.~\ref{fig:06}), well below the MPP-QTE transition
at $\Omega=\tilde{\Omega}_{2}\simeq2\Omega_{1}$, the magnetic ratios
$\mathcal{B}_{xy(zy)}$ have values in the range 0.5-1, making the
polarization of these modes distinct from transverse magnetic (which
have $\mathcal{B}_{xy(zy)}=0$). For $\Omega\gtrsim\Omega_{1}$, the
magnetic ratio $\mathcal{B}_{xy}$ decreases with increasing frequency/$q^{\prime}$,
until it reaches a minimum at $\Omega/\Omega_{1}\simeq1.73$ of about
$0.05$ (not shown). The other MPP branches display similar behavior:
a strong variation of polarization near at the transitions, but with
$\mathcal{B}_{xy(zy)}$ never reaching negligible values.

We have found no evidence for MPP modes with $\mathcal{B}_{xy(zy)}\approx0$
for other choices of LL occupancy and broadening, as well. In 2D electron
gases, the situation is very distinct, since large wave-vector modes
are essentially longitudinal.\cite{Chiu_Quinn_1974,Bardos_Frankel_1994}
A question that deserves further investigation is whether the effect
of finite $\boldsymbol{q}=q^{\prime}\boldsymbol{e}_{x}$ in $\sigma_{L}$
(and $\sigma_{H}$) can influence the solutions at large $q^{\prime}$.
A considerable renormalization of spectrum is expected for the largest
wave vectors presented in our plots, which are of the order of $l_{B}^{-1}$
(see, e.g.,~Fig.~\ref{fig:02}). The lack of longitudinal character
of MPP reported here may indeed result from an inadequacy of the optical
limit in describing large $q^{\prime}$ modes. On the other hand,
the system represented in Fig.~\ref{fig:04} displays modes with
smaller wave vectors {[}note that branches with $n>2$ have $q^{\prime}(\Omega)\lesssim l_{B}^{-1}${]},
making the optical limit less restrictive in this case. Nevertheless,
similar features are observed in this system, thus providing further
evidence for the generality of the phenomenons discussed in this paper,
at least for small values of $q^{\prime}l_{B}$ .

\subsection*{MPP wave localization}

\begin{figure}
\centering{}\includegraphics[clip,width=0.75\columnwidth]{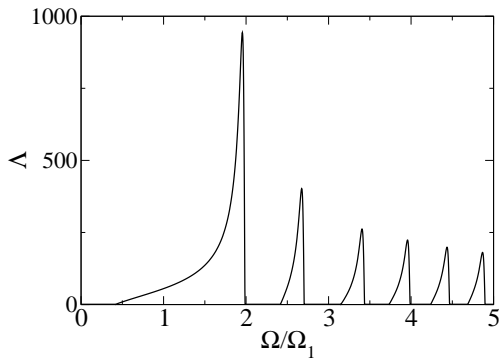}\caption{\label{fig:07}The wave localization ratio $\Lambda$ is plotted as
a function of the mode's frequency. Other parameters as in Fig.~\ref{fig:01}.}
\end{figure}

We briefly address the wave-localization characteristics of the MPP
waves reported here. It is a well-established fact that SPP in a metal
can have wavelengths considerably smaller than electromagnetic waves
of the same frequency in a dielectric.\cite{Barnes2010} In graphene
in zero field, this shrinkage effect is enhanced when compared to
conventional 2D electron gas SPP.\cite{plasmonics_graph_zerofield_Jablan,Koppens}
Figure~\ref{fig:07} shows the ratio of the wavelength in vacuum
to the MPP mode wavelength, $\Lambda=\lambda_{0}/\lambda$ (here,
$\lambda_{0}=2\pi c/\Omega$ and $\lambda=2\pi/q^{\prime}$). Near
the frequency resonant to the first interband transition at $\Omega\simeq2\Omega_{1}$(see
also Fig.~\ref{fig:01}), we obtain a large peak of about $\Lambda\sim10^{3}$,
a figure comparable to that obtained in zero field.\cite{plasmonics_graph_zerofield_Jablan}
The remaining MPP modes show peaks with $\Lambda\sim10^{2}$. QTE
solutions, on the other hand, have $\Lambda\simeq1$ regardless of
their frequency, a characteristic of transverse electric modes.

A simple formula for $\Lambda$, valid in the intraband frequency
region, can be derived from Eq.~(\ref{eq:omega_MPP_semiclassical}):
\begin{equation}
\Lambda\approx\frac{1}{2\alpha E_{F}}\left(\hbar\Omega-\frac{[\Delta(B)]^{2}}{\hbar\Omega}\right)\,,\label{eq:wave_loc}
\end{equation}
where $\alpha=e^{2}/(4\pi\epsilon\hbar c)$ denotes the effective
fine-structure constant and $\Delta(B)=\hbar ev_{F}^{2}B/E_{F}$ is
the cyclotron energy. The above expression predicts a decrease of
$\Lambda$ with the magnetic field, which is consistent with the exact
numerical results in Fig.~\ref{fig:07} (recall that $\Omega_{1}=\sqrt{2}v_{F}/l_{B}$).
In the limit of $B\rightarrow0$, Eq.~(\ref{eq:wave_loc}) reproduces
the result reported in Ref.~\cite{Koppens}.

\section{Outlook and Conclusions\label{sec:Outlook-and-Conclusions}}

We have computed the spectrum of electromagnetic modes supported by
a graphene interface in the presence of a quantizing magnetic field.
We have found a rich structure with extended crossovers between \emph{quasi}-transverse-electric
(QTE) and magnetoplasmon polariton (MPP) modes as a consequence of
characteristic Shubnikov\textendash{}de Haas oscillations in the magneto-optical
response of graphene. Analogously to the 2D Fermi gas, the dispersion
relation splits in a large number of branches.\cite{Chiu_Quinn_1974,Bardos_Frankel_1994}
Interband transitions between the bottom and top Dirac cones originate
terms in the conductivity with considerable spectral weight in the
semi-classical (low-frequency) region. As a consequence, the conventional
semi-classical 2D magnetoplasmon dispersion becomes restricted to
a narrow wave-vector/frequency interval. Our calculation within the
optical-limit approximation to the conductivity predicts that, unlike
2D Fermi gases, MPP modes with a given wave vector admit two possible
values of frequency in the same MPP branch. The consequences of a
non-zero wave vector $\boldsymbol{q}$ in the conductivity for the
QTE-MPP spectrum is a challenging question and deserves further investigation.

In summary, we have shown that a quantizing magnetic field changes
the conventional picture of electromagnetic modes in graphene.\cite{ZieglerPRL,plasmonics_graph_zerofield_Jablan}
In the LL regime, the mode spectrum splits into many branches, consisting
of magnetoplasmon polaritons (MPP) and \emph{quasi}-transverse-electric
(QTE) modes. At small frequencies, a QTE persists, even in the semi-classical
regime, as long as the cyclotronic frequency is larger than the electrons
relaxation's rate. Due to the high LL energy gaps, these effects should
be observable up to room temperature.

\section{Acknowledgements}

This work was supported by the NRF-CRP award \textquotedbl{}Novel
2D materials with tailored properties: beyond graphene\textquotedbl{}
(R-144-000-295-281).

\section{Appendix A: Mode Spectrum}

For our purposes it is sufficient to consider a single Fourier component
of the electromagnetic field in the following form 
\begin{equation}
\boldsymbol{E}_{m}=(E_{m,x},E_{m,y},E_{m,z})e^{i(q_{m}x-\omega t)}e^{-\kappa_{m}|z|}\,,\label{eq:ansatz}
\end{equation}
where $\kappa_{m}$ accounts for possible attenuation in the transverse
direction to graphene. The subscript $m=1(2)$ denotes the region
of space with $z>0$($z<0$). In this section, we consider the more
general case of graphene embedded in dielectric mediums with permittivities
$\epsilon_{1}$ and $\epsilon_{2}$. 

We require $\kappa_{m}^{\prime}\ge0$ and $q_{m}^{\prime\prime}/q_{m}^{\prime}\ge0$.
The former means that the wave can be confined to the $z$ plane and
must not diverge as $z\rightarrow\pm\infty$, whereas the second condition
allows wave attenuation in the $x$ direction. The formal divergence
at $x\rightarrow-\infty$ is a consequence of the beginning of the
perturbation at $t=-\infty$; for a detailed discussion on the built-in
divergence of leaky waves in the context of SPP see Ref.~\onlinecite{Glass}.

The macroscopic Maxwell equations imply the following relation between
the field amplitudes, 
\begin{equation}
i\omega\overrightarrow{\mathcal{F}_{m}}=M_{m}\overrightarrow{\mathcal{F}_{m}}\,,\label{eq:Maxwell_Relations}
\end{equation}
 where $\overrightarrow{\mathcal{F}_{m}}$ is the six-dimensional
vector of amplitudes $\overrightarrow{\mathcal{F}_{m}}=(\boldsymbol{E}_{m},\boldsymbol{B}_{m})$
and, 
\begin{equation}
M_{m}=\left[\begin{array}{cc}
0 & M_{m,R}\\
M_{m,L} & 0
\end{array}\right]\,,\label{eq:matrix_M}
\end{equation}
with 
\begin{align}
M_{m,R} & =\frac{1}{\epsilon_{m}\mu_{m}}\left[\begin{array}{ccc}
0 & (-1)^{m}\kappa_{m} & 0\\
(-1)^{m+1}\kappa_{m} & 0 & iq_{m}\\
0 & -iq_{m} & 0
\end{array}\right]\,,\label{eq:M_R}\\
M_{m,L} & =\left[\begin{array}{ccc}
0 & (-1)^{m+1}\kappa_{m} & 0\\
(-1)^{m}\kappa_{m} & 0 & -iq_{m}\\
0 & iq_{m} & 0
\end{array}\right]\,.\label{eq:M_L}
\end{align}
A straightforward consequence of Eq.~(\ref{eq:Maxwell_Relations})
is the well-known relation,
\begin{equation}
q_{m}^{2}=\kappa_{m}^{2}+\epsilon_{m}\mu_{m}\omega^{2}\,.\label{eq:general_relation}
\end{equation}
The mode spectrum for this problem is derived by imposing the boundary
conditions for the electromagnetic field at the interface $z=0$ and
making use of the relations between the field components {[}Eqs.~(\ref{eq:Maxwell_Relations})-(\ref{eq:M_L}){]}.
The continuity of the tangential (normal) component of the electric
field (magnetic induction) implies that $q_{1}=q_{2}\equiv q$, $E_{1,x(y)}=E_{2,x(y)}$
and $B_{1,z}=B_{2,z}$. The discontinuity of the tangential component
of the magnetic field yields 
\begin{eqnarray}
\frac{1}{\mu_{1}}B_{1,y}-\frac{1}{\mu_{2}}B_{2,y} & = & -\sigma_{xx}E_{1,x}-\sigma_{xy}E_{1,y}\,,\label{eq:condition3}\\
\frac{1}{\mu_{1}}B_{1,x}-\frac{1}{\mu_{2}}B_{2,x} & = & \sigma_{yx}E_{1,x}+\sigma_{yy}E_{1,y}\,,\label{eq:condition4}
\end{eqnarray}
where $\sigma_{\alpha\beta}\equiv\sigma_{\alpha\beta}(\boldsymbol{q},\omega)$
{[}$\alpha,\beta=x,y${]} denotes the dynamical conductivity of graphene,
\begin{equation}
\sigma_{\alpha\beta}(\boldsymbol{q},\omega)=\frac{J_{s,\alpha}(\boldsymbol{q},\omega)}{E_{s,\beta}(\boldsymbol{q},\omega)}\,,\label{eq:Ohm}
\end{equation}
and relates the Fourier transforms of the surface current, $J_{s,\alpha}(\boldsymbol{q},\omega)$,
and that of the electric field, $E_{s,\beta}(\boldsymbol{q},\omega)$,
at $z=0$. The use of the local limit of the conductivity $\sigma_{\alpha\beta}(\boldsymbol{q}=0,\omega)$
is justified whenever the wave vectors of interest are much smaller
than the inverse of typical length scales. In the presence of a quantizing
magnetic field, the local limit is justified for $|\boldsymbol{q}|\ll l_{B}^{-1}$,
where $l_{B}$ denotes the magnetic length (see Appendix B). 

Combining the above results, it is straightforward to obtain the general
dispersion relation
\begin{eqnarray}
\frac{i\omega}{\kappa_{1}}\epsilon_{1}\left[\left(1+\frac{\kappa_{1}}{\kappa_{2}}\frac{\epsilon_{2}}{\epsilon_{1}}\right)-\sigma_{L}\frac{\kappa_{1}}{i\omega\epsilon_{1}}\right]\times\nonumber \\
\times\left[\sigma_{L}-\left(\frac{\kappa_{1}}{\mu_{1}}+\frac{\kappa_{2}}{\mu_{2}}\right)\frac{1}{i\omega}\right] & = & \sigma_{H}^{2}\,.\label{eq:general_wave_spectrum}
\end{eqnarray}
In order to obtain the above form, we have invoked rotational symmetry
and used the notation employed in the main text: $\sigma_{L}\equiv\sigma_{xx}$
and $\sigma_{H}\equiv\sigma_{xy}$. Setting $\epsilon_{1}(\mu_{1})=\epsilon_{2}(\mu_{2})\equiv\epsilon(\mu)$
(and hence $\kappa_{1}=\kappa_{2}\equiv\kappa$) leads to the Eq.~(\ref{eq:fundamental_relation})
in the main text. We remark that the term with $\sigma_{H}$ is negligible
for most choices of parameters. We have verified that only near at
the QTE/MPP transitions, where $\sigma_{L}^{\prime\prime}\simeq0$,
the Hall conductivity $\sigma_{H}$ provides a (small) correction
to the spectrum.

\section{Appendix B: Magneto-Optical Conductivity of Graphene}

Within the Dirac-cone approximation,\cite{rmp} and modeling the effect
of disorder by an energy broadening function, the magneto-optical
conductivity of graphene at Fermi energy $E_{F}$ and temperature
$T$ assumes the simple form in the random phase approximation\cite{Sharapov,EOM}
\begin{align}
\sigma_{L(H)}(\omega,B) & =g_{s}g_{v}\times\frac{e^{2}}{4h}\times\nonumber \\
 & \times\sum_{n\neq m}\frac{\Xi_{L(H)}^{nm}}{i\Delta{}_{nm}}\frac{n_{F}(E_{n})-n_{F}(E_{m})}{\hbar\omega+\Delta{}_{nm}+i\Gamma_{nm}(\omega)}\,,\label{eq:conductivity}
\end{align}
where $g_{s(v)}=2$ is the spin (valley) degeneracy factor of graphene,
$n_{F}(E)=1/[1+e^{(E-E_{F})/k_{B}T}]$ stands for the Fermi distribution
function, $\Gamma_{nm}(\omega)$ is the LL broadening, $\Delta{}_{nm}=E_{n}-E_{m}$,
with LL energies $E_{n}$ given by 
\begin{equation}
E_{n}=\textrm{sign}(n)[\hbar v_{F}/l_{B}]\sqrt{2|n|}\,,\label{eq:Landau_Levels}
\end{equation}
with $l_{B}$ denoting the magnetic length, $l_{B}\equiv\sqrt{\hbar/(eB)}$,
$v_{F}\simeq10^{6}$m/s is the Fermi velocity, and 
\begin{eqnarray}
\Xi_{L}^{nm} & = & \frac{\hbar^{2}v_{F}^{2}}{l_{B}^{2}}(1+\delta_{m,0}+\delta_{n,0})\delta_{|m|-|n|,\pm1}\,,\label{eq:tensor_diagonal}\\
\Xi_{H}^{nm} & = & i\Xi_{L}^{nm}(\delta_{|m|,|n|-1}-\delta_{|m|-1,|n|})\,.\label{eq:tensor_offdiagonal}
\end{eqnarray}
The use of the low-energy (Dirac-cone approximation) theory to compute
$\sigma_{L(H)}$ assumes an infinite sea of negative energy states,
and thus requires a cutoff $|n|,|m|\le N_{\textrm{cut}}$ in Eq.~(\ref{eq:conductivity}).
The respective cutoff energy $E_{N_{\textrm{cut}}}$ is of the order
of graphene's bandwidth. Results are largely insensitive to the precise
value chosen for the cutoff; in our numerical calculations we have
considered $E_{N_{\textrm{cut}}}=3$~eV.

The magneto-optical conductivity of graphene has two types of terms:
i) intraband contributions corresponding to transitions within the
same Dirac cone (i.e.,~$n=m\pm1$), and ii) interband transitions
that couple the valence and conduction Dirac cones (i.e.,~$n=-m\pm1$).
Transitions involving the zero energy LL (e.g., $n=0\rightarrow n=1$)
need to be considered separately because this LL state contains electrons
and holes. Here, for convenience, we classify the transitions involving
the zero energy LL as interband-like. 

The general expression Eq.~(\ref{eq:conductivity}) can be put into
a more useful form by separating interband and intraband contributions
(with the proviso made in the previous paragraph). For the sake of
simplicity, we assume $T=0$ and $E_{F}\ge0$ {[}the conductivity
for holes can be obtained using the symmetry relations: $\sigma_{L}(-E_{F})=\sigma_{L}(E_{F})$
and $\sigma_{H}(-E_{F})=-\sigma_{H}(E_{F})${]}. We denote the number
of occupied electron LLs by $N_{F}$, that is, $N_{F}=\textrm{int}\left[(E_{F}/E_{1})^{2}\right]\ge0$.
Intraband transitions ($n=N_{F}\rightarrow n=N_{F}+1$) involve an
energy difference of $\Delta_{\textrm{intra}}=\sqrt{2}\hbar v_{F}/l_{B}(\sqrt{N_{F}+1}-\sqrt{N_{F}})$.
Its contribution to the conductivity dominates at small frequencies
where most of the spectral weight is concentrated around $\omega=\Delta_{\textrm{intra}}/\hbar$.
The intraband (semi-classical) conductivity therefore consists of
a single term in the summation Eq.~(\ref{eq:conductivity}), reading
\begin{align}
\sigma_{L}^{(\textrm{intra})} & =\frac{2e^{2}}{h}\frac{\hbar^{2}v_{F}^{2}}{l_{B}^{2}\Gamma\Delta_{\textrm{intra}}}\frac{1-i\hbar\omega/\Gamma}{(1-i\hbar\omega/\Gamma)^{2}+\Delta_{\textrm{intra}}^{2}/\Gamma^{2}}\,,\label{eq:sigma_L_intra}\\
\sigma_{H}^{(\textrm{intra})} & =-\frac{2e^{2}}{h}\frac{\hbar^{2}v_{F}^{2}}{l_{B}^{2}\Gamma^{2}}\frac{1}{(1-i\hbar\omega/\Gamma)^{2}+\Delta_{\textrm{intra}}^{2}/\Gamma^{2}}\,.\label{eq:sigma_H_intra}
\end{align}
According to our classification the latter equations are valid for
$N_{F}\neq0,$ otherwise there is no intraband contribution. Note
that for high Fermi energy/low magnetic field, one recovers the familiar
semi-classical Drude conductivity {[}Eqs.~(\ref{eq:sigma_xx_semiclass})
and (\ref{eq:sigma_xy_semiclass}){]}, since the cyclotronic gap $\Delta_{\textrm{intra}}$
equals the cyclotronic energy $\hbar\omega_{\textrm{c}}$ when many
levels are occupied, $N_{F}\gg1$.\cite{EOM} 

Interband terms dominate at frequencies close or above the interband
threshold, $\omega=2E_{F}/\hbar$. These transitions involve the energy
difference energy $\Delta_{n}\equiv\sqrt{2}\hbar v_{F}/l_{B}(\sqrt{n+1}+\sqrt{n})$,
with $n\ge N_{F}$. Its contribution to the magneto-optical conductivity
reads as
\begin{align}
\sigma_{L}^{(\textrm{inter})} & = & \frac{2e^{2}}{h}\frac{\hbar^{2}v_{F}^{2}}{l_{B}^{2}}\sum_{n=N_{F}}^{N_{\textrm{cut}}}(1+\delta_{n,0})(2-\delta_{n,N_{F}})\times\nonumber \\
 &  & \times\frac{1}{\Gamma\Delta_{n}}\frac{1-i\hbar\omega/\Gamma}{(1-i\hbar\omega/\Gamma)^{2}+\Delta_{n}^{2}/\Gamma^{2}}\,,\label{eq:sigma_L_inter}\\
\sigma_{H}^{(\textrm{inter})} & = & -\frac{2e^{2}}{h}\frac{\hbar^{2}v_{F}^{2}}{l_{B}^{2}\Gamma^{2}}\frac{1+\delta_{N_{F},0}}{(1-i\hbar\omega/\Gamma)^{2}+\Delta_{N_{F}}^{2}/\Gamma^{2}}\,.\label{eq:sigma_H_inter}
\end{align}

\section{Appendix C: Semi-Classical Dispersion}

We now derive an approximate formula for the semi-classical MPP dispersion.
The first step is to solve Eq.~(\ref{eq:fundamental_relation}) for
$\kappa(\Omega)$; we obtain
\begin{equation}
\kappa(\Omega)=\frac{X(\Omega)\pm\sqrt{4f(\Omega)g(\Omega)+X(\Omega)^{2}}}{2f(\Omega)}\,,\label{eq:sol_mag1}
\end{equation}
with the notation,
\begin{eqnarray}
f(\Omega) & = & \frac{i}{2\Omega\epsilon}\sigma_{L}(\Omega,B)\,,\label{eq:aux_sol_mag_1}\\
g(\Omega) & = & \frac{i\Omega\mu}{2}\sigma_{L}(\Omega,B)\,,\label{eq:aux_sol_mag_2}\\
X(\Omega) & = & -1+f(\Omega)g(\Omega)-\frac{\mu}{4\epsilon}\sigma_{H}^{2}(\Omega,B)\,.\label{eq:aux_sol_mag_3}
\end{eqnarray}
The sign in the numerator in Eq.~(\ref{eq:sol_mag1}) must be chosen
according to the requirements necessary to obtain a physical solution
(Appendix A). The complex wave vector $q(\omega)$ follows from Eq.~(\ref{eq:general_relation}),
\begin{equation}
q(\Omega)=\sqrt{\kappa(\Omega)^{2}+\epsilon_{\textrm{r}}\Omega^{2}/c^{2}}\,,\label{eq:q(omega)}
\end{equation}
where $\epsilon_{\textrm{r}}$ denotes the relative permittivity of
the dielectric medium surrounding graphene.\cite{comment_2dielectrics}
In order to proceed, we neglect the effect of interband transitions
and approximate Eqs.~(\ref{eq:sigma_L_intra}) and (\ref{eq:sigma_H_intra})
by their semi-classical analog\cite{EOM}
\begin{align}
\sigma_{L} & =\frac{e^{2}}{h}\frac{2|E_{F}|}{\Gamma}\frac{1-i\hbar\Omega/\Gamma}{(1-i\hbar\Omega/\Gamma)^{2}+\Delta^{2}/\Gamma^{2}}\,,\label{eq:sigma_xx_semiclass}\\
\sigma_{H} & =-\frac{e^{2}}{h}\frac{2E_{F}}{\Gamma}\frac{\Delta/\Gamma}{(1-i\hbar\Omega/\Gamma)^{2}+\Delta^{2}/\Gamma^{2}}\,,\label{eq:sigma_xy_semiclass}
\end{align}
where $\Delta=\hbar ev_{F}^{2}B/|E_{F}|$ is the intraband cyclotron
gap. The semi-classical expressions have the advantage of simplifying
the notation and introducing the cyclotron energy which is more used
in the literature (albeit less accurate than the intraband gap, $\Delta_{\textrm{intra}}$;
see Appendix B). The crucial point to derive a compact expression
for the dispersion relation is to note that for typical frequencies
$\sim$THz and $\Gamma\sim0.01$~eV, we have $f\times g\ll X^{2}$,
and therefore the expression for $\kappa(\Omega)$ can be approximated
by 
\begin{equation}
\kappa(\Omega)\simeq\frac{X(\Omega)}{f(\Omega)}\,,\label{eq:sol_kay_approx}
\end{equation}
where we have chosen the appropriate sign in Eq.~(\ref{eq:sol_mag1}).
{[}Note that the other solution has $\kappa(\Omega)\simeq0$ and would
correspond to a QTE mode.{]} Raising both sides of Eq.~(\ref{eq:sol_kay_approx})
to the power of two, substituting the conductivity tensor components
{[}Eqs.~(\ref{eq:sigma_xx_semiclass}) and (\ref{eq:sigma_xy_semiclass}){]}
into $X(\Omega)$, and employing a series expansion for small $\Delta/\hbar\Omega$
and $\Gamma/\hbar\Omega$, we obtain, 
\begin{equation}
q^{\prime2}+q^{\prime\prime2}\simeq\epsilon_{\textrm{r}}\frac{\Omega^{2}}{c^{2}}+\left(\frac{2\pi\epsilon\hbar\Omega}{E_{F}e^{2}}\right)^{2}\left(\hbar^{2}\Omega^{2}-2\Delta^{2}\right)\,,\label{eq:aux_spect_mag}
\end{equation}
where we have kept the terms up to second order in the small parameters,
and assumed $\hbar\Omega/E_{F}\gg g_{0}\sqrt{\mu/\epsilon}\simeq0.1$
(with $g_{0}\equiv2e^{2}/h$ denoting the quantum of conductance).
We remark that these approximations are consistent with the small-wavelength
limit, for which $\hbar\Omega$ is typically larger than other energy
scales. Assuming negligible damping $q^{\prime\prime}\ll q^{\prime}$,
more precisely, requiring
\begin{equation}
\Gamma\ll\frac{E_{F}^{2}e^{4}q^{\prime2}}{\epsilon^{2}\hbar^{3}\Omega^{3}}\,,\label{eq:small_damping}
\end{equation}
and taking $q^{\prime}\gg\Omega/c$ (non-retarded regime) in Eq.~(\ref{eq:aux_spect_mag}),
we arrive at the final result,
\begin{equation}
q^{\prime}\approx\frac{2\pi\epsilon\hbar\Omega}{|E_{F}|e^{2}}\sqrt{\hbar^{2}\Omega^{2}-2\Delta^{2}}\,.\label{eq:relation_spectrum_in_field_approx}
\end{equation}
The magneto-plasmon spectrum {[}Eq.~(\ref{eq:omega_MPP_semiclassical}){]}
follows immediately by expanding the latter expression in the small
parameter $\Delta/\hbar\Omega$.


\begin{thebibliography}{10}
\bibitem{Ebbesen1998}T. W. Ebbesen, H. J. Lezec, H. F. Ghaemi, T.
Thio, and P. A. Wolf, Nature \textbf{391}, 667 (1998).

\bibitem{PlasmonBook}S. A. Maier, Plasmonics: Fundamentals and Applications,
Springer (2007).

\bibitem{Ashkan}A. Vakil, and N. Engheta, Science \textbf{332}, 1291
(2011).

\bibitem{Schedin2010}F. Schedin, E. Lidorikis, A. Lombardo, V. G.
Kravets, A. K. Geim, A. N. Grigorenko, K. S. Novoselov, and A. C.
Ferrari, ACSNano \textbf{4}, 5617 (2010).

\bibitem{EchtermeyerPhoto}T. J. Echtermeyer, L. Britnell, P. K. Jasnos,
A. Lombardo, R. V. Gorbachev, A. N. Grigorenko, A. K. Geim, A. C.
Ferrari, and K. S. Novoselov, Nature Communications \textbf{2}, 458
(2011).

\bibitem{LongJuPlasmonics}L. Ju, B. Geng, J. Horng, C. Girit, M.
C. Martin, Z. Hao, H. A. Bechtel, X. Liang, A. Zettl, Y. R. Shen,
and F. Wang, Nature Nanotechnology \textbf{6}, 630 (2011).

\bibitem{dirac-plamons}Z. Fei, G. O. Andreev, W. Bao, L. M. Zhang,
A. S. McLeod, C. Wang, M. K. Stewart, Z. Zhao, G. Dominguez, M. Thiemens,
M. M. Fogler, M. J. Taube, A. H. Castro-Neto, C. N. Lau, F. Keilmann,
and D. N. Basov, Nano Lett. \textbf{11}, 4701 (2011).

\bibitem{rmp}A. H. Castro Neto, F. Guinea, N. M. R. Peres, K. S.
Novoselov, and A. K. Geim, Rev. Mod. Phys\textbf{ 81}, 109 (2009).

\bibitem{YuliyEPL}Yu. V. Bludov, M. I. Vasilevskiy, and N. M. R.
Peres, EPL \textbf{92}, 68001 (2010).

\bibitem{plasmonics_graph_zerofield_Jablan}M. Jablan, H. Buljan,
and M. Solja\v{c}i\'{c}, Phys. Rev. B \textbf{80}, 245435 (2009).

\bibitem{Koppens}F. H. L. Koppens, D. E. Chang, and F. J. G. de Abajo,
Nano Lett. \textbf{11}, 3370 (2011).

\bibitem{Chiu_Quinn_1974}K. W. Chiu, and J. J. Quinn, Phys. Rev.
B \textbf{9}, 4724 (1974).

\bibitem{exp2dmspp}I. V. Kukushkin, V. M. Muravev, J. H. Smet, M.
Hauser, W. Dietsche, and K. von Klitzing, Phys. Rev. B \textbf{73},
113310 (2006).

\bibitem{Bychkov2008}Y. A. Bychkov, and G. Martinez, Phys. Rev. B
\textbf{77}, 125417 (2008). 

\bibitem{Berman2008}O. L. Berman, G. Gumbs, and Y. E. Lozovik, Phys.
Rev. B \textbf{78}, 085401 (2008). 

\bibitem{comment_2dielectrics}Usually experiments are done with graphene
on top of a substrate such as SiO$_{2}$ (see e.g.,~Ref.~\cite{dirac-plamons}).
The presence of two distinct dielectrics can be accounted for letting
$\epsilon\rightarrow(\epsilon_{1}+\epsilon_{2})/2$, where $\epsilon_{1(2)}$
concerns with the top (bottom) dielectric. This prescription is accurate
in the non-retarted regime $q^{\prime}\gg\Omega/c$; for the exact
dispersion relation we refer to Appendix~A.

\bibitem{Barnes2010}W. L. Barnes, A. Dereux, and T. W. Ebbesen, Nature
(London) \textbf{424}, 824 (2003).

\bibitem{ZieglerPRL}S. A. Mikhailov, and K. Ziegler, Phys. Rev. Lett.
\textbf{99}, 016803 (2007).

\bibitem{TEpolarizer}Q. Bao, H. Zhang, B. Wang, Z. Ni, C. H. Y. X.
Lim, Y. Wang, D. Y. Tang, and K. P. Loh, Nature Photonics \textbf{5},
411(2011).

\bibitem{pump_probe_epitaxial}J. M. Dawlaty, S. Shivaraman, M. Chandrashekhar,
F. Rana, and M. G. Spencer, Appl. Phys. Lett. \textbf{92}, 042116
(2008).

\bibitem{pump_probe_exfoliated}M. Breusing, S. Kuehn, T. Winzer,
E. Mali\'{c}, F. Milde, N. Severin, J. P. Rabe, C. Ropers, A. Knorr,
and T. Elsaesser, Phys. Rev. B \textbf{83}, 153410 (2011).

\bibitem{Horgn2011}J. Horng, C.-F. Chen, B. Geng, C. Girit, Y. Zhang,
Z. Hao, H. A. Bechtel, M. Martin, A. Zettl, M. F. Crommie, Y. R. Shen,
and F. Wang, Phys. Rev. B \textbf{83}, 165113 (2011).

\bibitem{Carbotte_Phonons}A. Pound, J. P. Carbotte, and E. J. Nicol,
EPL \textbf{94}, 57006 (2011).

\bibitem{Carbotte_Phonons_2}A. Pound, J. P. Carbotte, and E. J. Nicol,
Phys. Rev. B \textbf{85}, 125422 (2012).

\bibitem{Sharapov}V.P. Gusynin, S.G. Sharapov, and J.P. Carbotte,
J. Phys.: Condens. Matter \textbf{19}, 026222 (2007).

\bibitem{EOM}A. Ferreira, J. Viana-Gomes, Yu. V. Bludov, V. Pereira,
N. M. R. Peres, A. H. Castro Neto,\emph{ }Phys. Rev. B \textbf{84},
235410 (2011).

\bibitem{Goerbig}R. Roldán, J.-N. Fuchs, and M. O. Goerbig, Phys.
Rev. B \textbf{80}, 085408 (2009). 

\bibitem{Bardos_Frankel_1994}D. C. Bardos, and N. E. Frankel, Phys.
Rev. B \textbf{49}, 4096 (1994).

\bibitem{comment2}In deriving this expression we have approximated
the conductivity in the intraband region by its semi-classical value.
The latter is consistent since i) the intraband gap $\hbar\Omega_{1}(\sqrt{N_{F}+1}-\sqrt{N_{F}})$
is well-approximated by the cyclotron energy $\hbar\omega_{\textrm{c}}$
in samples with $N_{F}\ge1$ (see Appendix B), and ii) the spectral
weight in this region coming from interband transitions is negligible
for $\Gamma\ll E_{1}$. We have also neglected the contribution from
the Hall condutivity to the dispersion relation (see Appendix A for
more details).

\bibitem{Jackson}J. D. Jackson, Classical Electrodynamics (Wiley,
New York, 1999).

\bibitem{Stern_Polarizability2D}F. Stern, Phys. Rev. Lett. \textbf{18},
546 (1967).

\bibitem{Shung_Graphite}K. W. K. Shung, Phys. Rev. B \textbf{34},
979 (1986).

\bibitem{Glass}N. E. Glass, M. Weber, and D. L. Mills, Phys. Rev.
B \textbf{29}, 6548 (1984). \end{thebibliography}
\end{document}